\documentclass[preprint,superscriptaddress]{revtex4-2}

\usepackage{graphicx}% Include figure files
\usepackage{dcolumn}% Align table columns on decimal point
\usepackage{bm}% bold math
\usepackage{natbib}
\usepackage{placeins}
\usepackage{amsmath}
\usepackage{amssymb}
\usepackage{graphicx}
\usepackage{epstopdf}
\usepackage{tikz}
\usepackage{multirow}
\usepackage{tabularx}
\usepackage{array}

\begin{document}

\renewcommand\tabularxcolumn[1]{m{#1}}% for vertical centering text in X column

\newcommand{\twopartdef}[6]
{
	\left\{
	\begin{array}{lll}
		#1 & \quad \mbox{for } #2 \\
		#3 & \quad \mbox{for } #4 
	\end{array}
	\right.
}

\title{Rate and memory effects in bifurcation-induced tipping} 
\date{\today}

\author{Julia Cantis\'{a}n }
\affiliation{Nonlinear Dynamics, Chaos and Complex Systems Group, Departamento de F\'{i}sica, Universidad Rey Juan Carlos \\ Tulip\'{a}n s/n, 28933 M\'{o}stoles, Madrid, Spain}
\author{Serhiy Yanchuk}
\affiliation{Department of Mathematics, Humboldt University Berlin, 12489, Berlin, Germany}
\affiliation{Potsdam Institute for Climate Impact Research, 14473, Potsdam, Germany}
\author{Jes\'{u}s M. Seoane}
\affiliation{Nonlinear Dynamics, Chaos and Complex Systems Group, Departamento de F\'{i}sica, Universidad Rey Juan Carlos \\ Tulip\'{a}n s/n, 28933 M\'{o}stoles, Madrid, Spain}
\author{Miguel A.F. Sanju\'{a}n}
\affiliation{Nonlinear Dynamics, Chaos and Complex Systems Group, Departamento de F\'{i}sica, Universidad Rey Juan Carlos \\ Tulip\'{a}n s/n, 28933 M\'{o}stoles, Madrid, Spain}
\affiliation{Department of Applied Informatics, Kaunas University of Technology \\ Studentu 50-415, Kaunas LT-51368, Lithuania}
\author{Jürgen Kurths}
\affiliation{Department of Physics, Humboldt University Berlin, 12489, Berlin, Germany}
\affiliation{Potsdam Institute for Climate Impact Research, 14473, Potsdam, Germany}

\begin{abstract}
A variation in the environment of a system, such as the temperature, the concentration of a chemical solution or the appearance of a magnetic field, may lead to a drift in one of the parameters. If the parameter crosses a bifurcation point, the system can tip from one attractor to another (bifurcation-induced tipping). Typically, this stability exchange occurs at a parameter value beyond the bifurcation value. This is what we call here the stability exchange shift. We study systematically how the shift is affected by the initial  parameter value and its change rate. To that end, we present numerical and analytical results for different types of bifurcations and different paradigmatic systems. Finally, we deduce the scaling laws governing this phenomenon. We show that increasing the change rate and starting the drift further from the bifurcation can delay the tipping process. Furthermore, if the change rate is sufficiently small, the shift becomes independent of the initial condition (no memory) and the shift tends to zero as the square root of the change rate. Thus, the bifurcation diagram for the system with fixed parameters is recovered. 
\end{abstract}

                            
\maketitle
\newpage

\section{Introduction} \label{Section_1}

The behavior of a dynamical system can be classified depending on the value of its parameters. Usually, this is represented by bifurcation diagrams, where the attractors are depicted for each region of parameters. For a fixed set of parameters and equations, the system remains in one of the corresponding attractors, either a fixed point, a limit cycle, a chaotic attractor, etc. However, in nature, this equilibrium is usually perturbed by the system's environment. Real-world systems are open and external perturbations can be modelled as a time variation of the parameters. A change in the parameters can cause the system to tip from one attractor to another, leading to a (sometimes drastic) regime shift. There are many examples of this sudden transitions in different fields such as ecology \cite{Hastings2018} or climate dynamics \cite{Lenton2008}. Furthermore, this phenomenon may be desirable and one can then design an input that causes the system to tip. For instance, this is the case of experiments that involve transitions through critical temperatures and yield phase changes \cite{Majumdar2013}. 

There are many different ways in which the variation of a parameter can lead to a regime shift. Following \cite{Ashwin2012a}, they can be classified in three categories: noise-induced tipping (N-tipping), rate-induced tipping (R-tipping) and bifurcation-induced tipping (B-tipping). In the case of N-tipping, a parameter suffers fluctuations that force the system to abandon the attractor and tip to another one. For R-tipping \cite{Wieczorek2011}, the rate at which a parameter drifts is sufficiently fast that the system fails to track the current attractor. Finally, B-tipping occurs when the parameter crosses a bifurcation, forcing the system to tip. 

Here, we deal with parameter drifts that lead to B-tipping, also called dynamic bifurcations \cite{Benoit1991} or slow passage through a bifurcation \cite{Baer1989} in previous literature. This phenomenon has been studied from different perspectives, exploring concepts such as tipping probability in multistable systems, scenario-dependent basins or early-warning indicators \cite{Neishtadt2009,Kaszas2019b,Ashwin2017,Song2021,Berglund1999a,Berglund2002}. We focus on describing in a systematic way the process of tipping, specifically the shift phenomenon \cite{Neishtadt1987}. It has been observed that in systems with a parameter drift the tipping appears for a value of the parameter $p$ different from the bifurcation value $ p_{b} $. We call this value the critical value $ p_{cr} $, which is $ p_{cr}>p_{b} $ (with  increasing $p$). The shift phenomenon has also been referred to as delay phenomenon, emphasizing the temporal delay in the tipping, $ t_{cr}>t_{b} $, rather than the shift in the parameter. Both terms are equivalent as the parameter depends on time. In Ref.~\cite{Hughes2013}, the time between $p_{cr}$ and $p_{b}$ is called borrowed time, to point out that in this time window the attractor is no longer stable, although the system is still tracking it, and it is still possible to avoid the tipping by reversing the drift.

The value of the shift $ p_{cr}-p_{0} $ is scenario-dependent and it may vary with the parameter change rate $ \varepsilon$. We call this the rate effect. This effect has been recently explored for complex systems such as maps with chaotic attractors \cite{Maslennikov2013a,Maslennikov2018a}, flows with chaotic attractors \cite{Cantisan2021} and time-delayed systems \cite{Cantisan2021a}. Another interesting variable that affects the value of $p_{cr}$ is the initial value of the parameter, $ p_{0} $. This implies that the system has memory and that the distance to the bifurcation is relevant even in the case when the equilibrium that looses stability is parameter independent. The memory effect has been less studied in \cite{Baer1989, Su2001} and it usually does not take into account the rate effect and their interplay.

In this paper, we aim at a systematic study of the shift phenomenon that appears in B-tipping. We will show that the system's dynamics is naturally split into two regimes: deterministic and non-deterministic, depending on whether (numerical) noise plays an ''essential'' role or not. The non-deterministic case means that the system's trajectory reaches the neighborhood of an attractor that is smaller than the threshold of the numerical noise. This is a typical situation occuring for a small change rate $\varepsilon$, as the local convergence to an attractor is exponential outside of the bifurcation. 
Also, realistic systems are always subject to some level of noise and experimental setups have a finite precision. 
In the deterministic regime, the numerical threshold is not reached. This is usually related to faster drifts when the system has no time to relax and come that closer to an attractor. We explore the rate and memory effects in the two regimes for different paradigmatic systems and different bifurcations. We obtain the analytical scaling laws that govern these processes and compare them with the numerical results.

The article is organized as follows. In Section \ref{Section_2}, we introduce the systems and bifurcations and the numerical techniques used to analyze them. In Section \ref{Section_3}, we present the results for the deterministic regime, whereas in Section \ref{Section_4} we present the ones dealing with the non-deterministic regime. The rate and memory effects are explored in the two regimes and both numerical and analytical calculations are presented. We provide concluding remarks at the end.

\section{Systems and bifurcations} \label{Section_2}

We choose paradigmatic examples of dynamical systems with different types of bifurcations as our aim is to study common responses to a changing parameter that crosses a bifurcation. 
To that end, we proceed in the same way for every system and compare the behavior in each case. We choose one of the parameters and we make it evolve linearly with time at a small but non-negligible rate $\varepsilon$. Depending on the natural timescale of the system, we take $\varepsilon$ in the range of $10^{-5}-10^{-2}$. The parameter sweep is set then following:
\begin{equation}
	p(t) = \twopartdef { p_{0} } {t<t_{1}} {p_{0}+\varepsilon \cdot (t-t_{1})} { t_{1}<t<t_{2}} 
 \label{eq_parameter}
	\quad \quad \quad \quad \quad \quad
	\begin{tikzpicture}[baseline]
		\draw[ultra thick] (0,-1) -- (-1.5,-1)  node[anchor=east] {$ p_{0} $};
		\draw[ultra thick] (0,-1) -- (1.5,1);
		\draw[ dotted] (0,1.5) -- (0,-1.5)  node[anchor=north] {$ t_{1} $};
		\draw[ dotted] (1.5,1.5) -- (1.5,-1.5) node[anchor=north] {$ t_{2} $};
	\end{tikzpicture}
 \label{eq_parameter}
\end{equation}
where $ t_{1} $ is the time for which the parameter drift starts and $ t_{2} $ when it ends. This way, we let the system evolve to its steady state, before the shift in the parameter starts. At some point between $t_{1}$ and $t_{2}$, the parameter crosses a bifurcation and tipping is expected to be observed with some delay. 

Specifically, we study: the pitchfork and subcritical Hopf bifurcations in the Lorenz system, the supercritical Hopf bifurcation in the FitzHugh-Nagumo model, and the period-doubling bifurcation in the Rössler system. By analyzing these systems, we cover some common bifurcations, including the period-doubling, which can lead to chaos. The following ODE solvers are used for the numerical integration:  ode78 (Runge-Kutta 8(7)) and ode15s (variable-step, variable-order (VSVO) solver). The latter method seems to capture the repulsion from an unstable attractor in a better way, but both of them yield quantatively the same results.

\subsection{Lorenz system}
The Lorenz system \cite{Lorenz1963} was proposed by Edward Lorenz as a simple model of convection dynamics in the atmosphere (Rayleigh-Bénard convection). The equations read as follows
\begin{equation}
	\begin{aligned}
		&\dot{x}=-\sigma x + \sigma y, \\
		&\dot{y}=r x -y -xz,\\
		&\dot{z}=-\beta z +xy,
	\end{aligned}
	\label{Lorenz}
\end{equation}
where $ \sigma $, $ \beta $, and $ r $ are the system parameters. In the context of convection dynamics, $ \sigma $ is the Prandtl number and is characteristic of the fluid, $ \beta $ depends on the geometry of the container, and $ r $ is the Rayleigh number that accounts for the temperature gradient. In this context, $ x $ represents the rotation frequency of convection rolls, while $ y $ and $ z $ correspond to variables associated to the temperature field. We fix the classical parameter values of $\sigma=10$, $ \beta=8/3 $, and consider $r$ as a time-depending parameter $r(t)$, following Eq.~(\ref{eq_parameter}). 

In the frozen-in system, where $r$ is a fixed parameter, we find successive bifurcations, see \cite{Cantisan2021} for their brief description. Here, we focus on the pitchfork and subcritical Hopf bifurcations that occur at $r=1$ and $r=24.74$, respectively. The rate effect for the heteroclinic bifurcation was analyzed previously in \cite{Cantisan2021}.
In the pitchfork bifurcation, the fixed point in the origin looses stability and two other fixed points are created:
\begin{equation}
   C^{\pm} =  (\pm \sqrt{\beta(r-1)}, \pm \sqrt{\beta(r-1)}, r-1 ),
   \label{eq_fixedpoint_Lorenz}
\end{equation}
while in the subcritical Hopf bifurcation, $C^{\pm}$ loose stability and a chaotic attractor, born at $r=24.06$, remains as the global attractor.

\begin{figure}[h]
	\begin{center}		
		\includegraphics[width=0.45\textwidth]{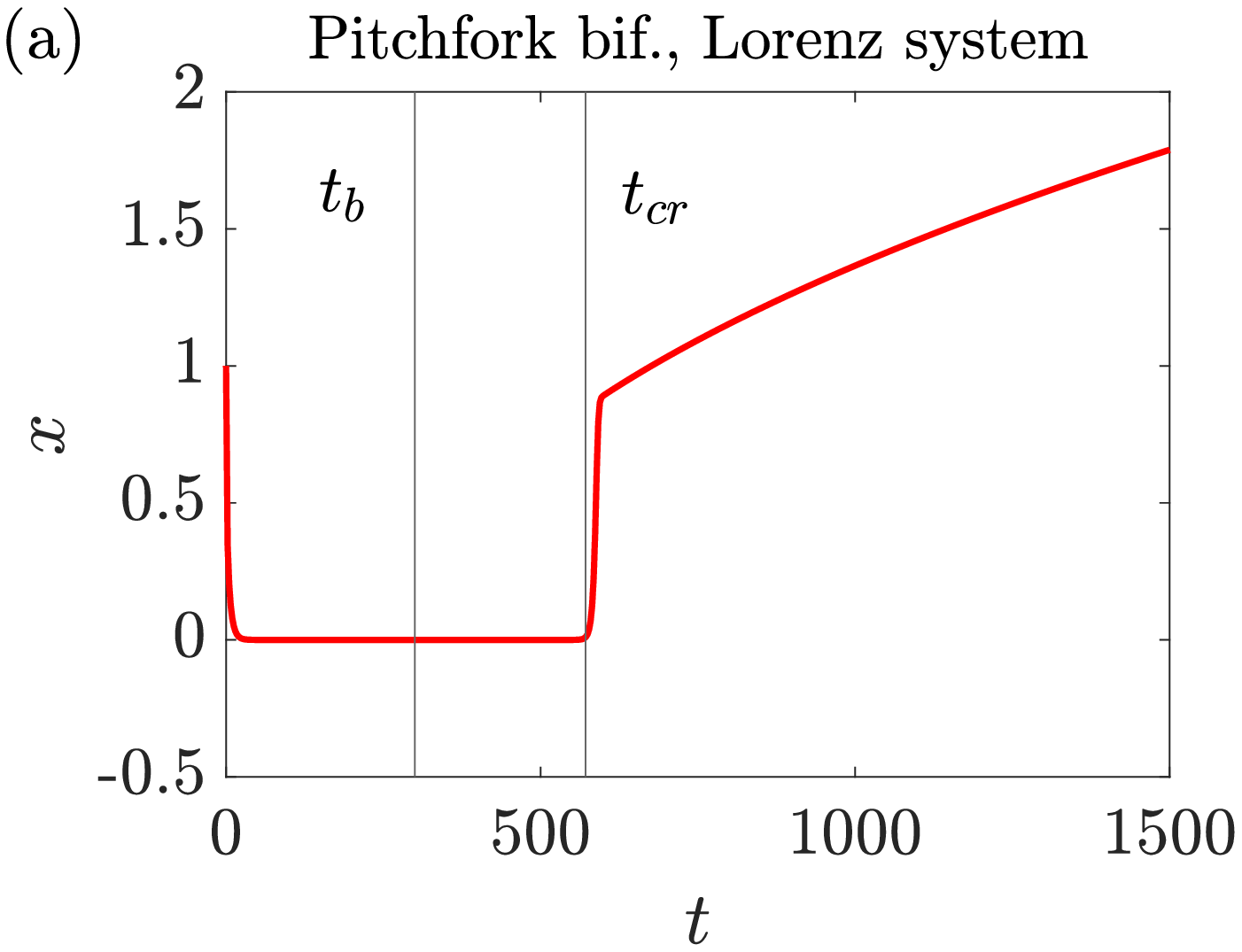}
		\includegraphics[width=0.45\textwidth]{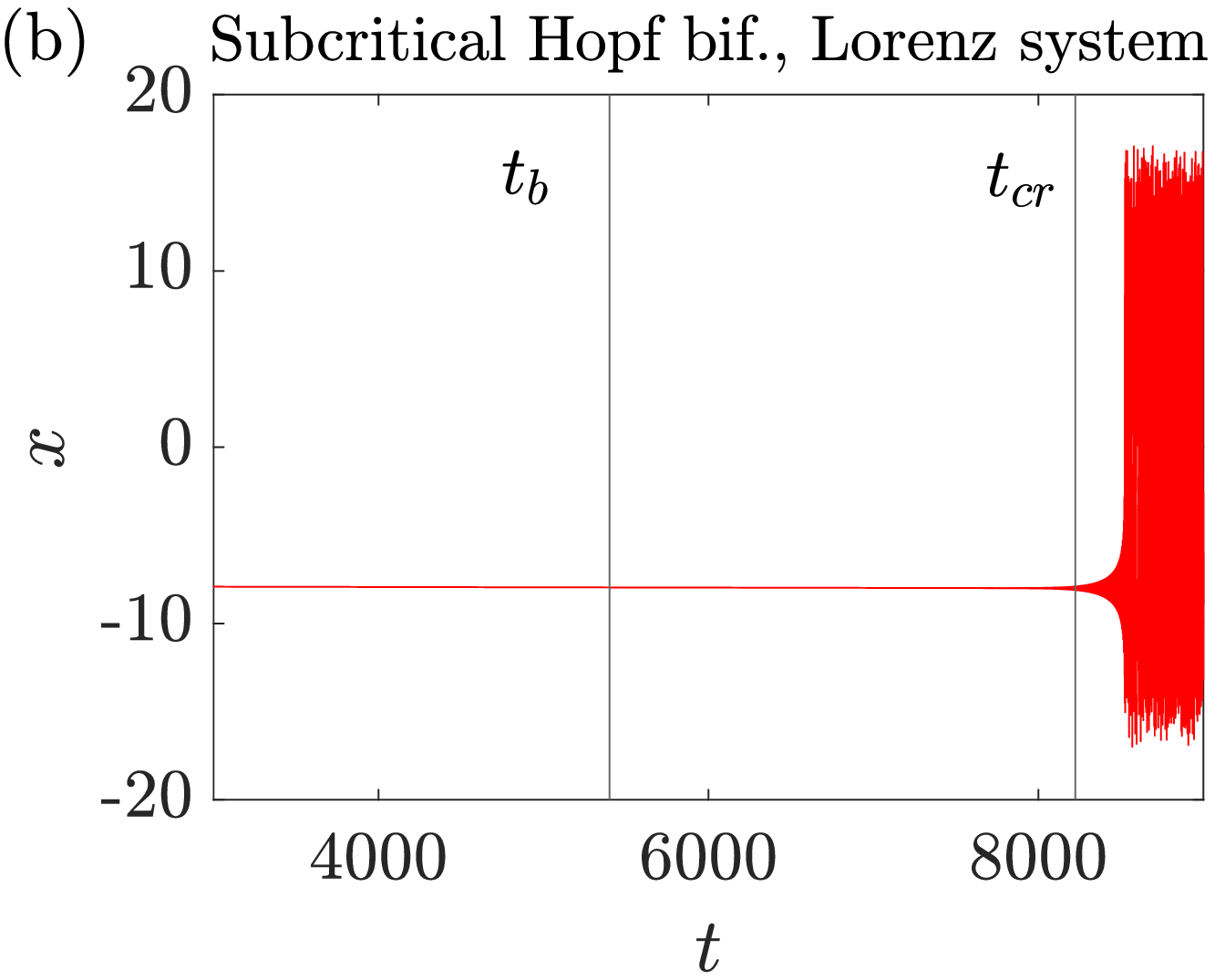}
		\includegraphics[width=0.45\textwidth]{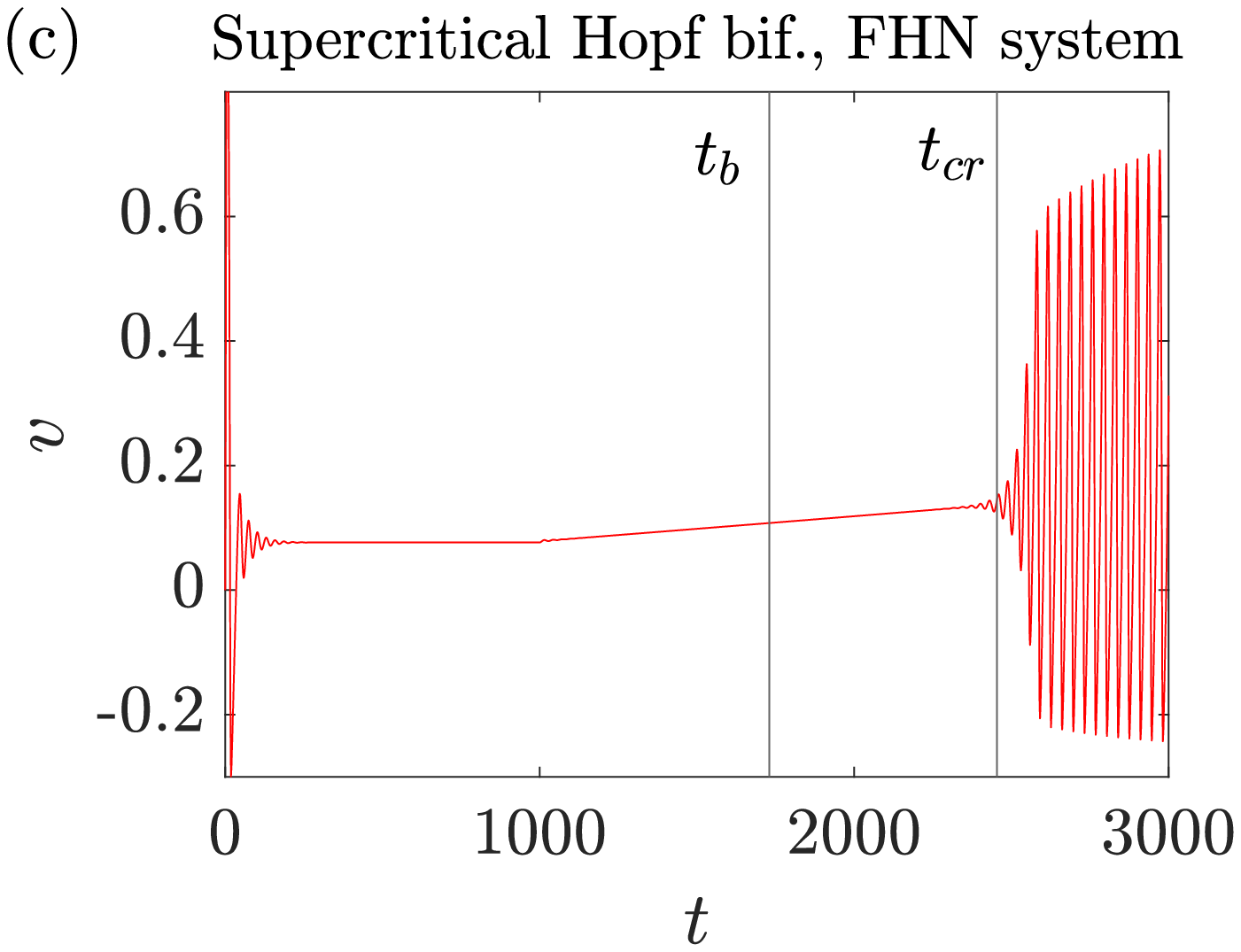}
		\includegraphics[width=0.45\textwidth]{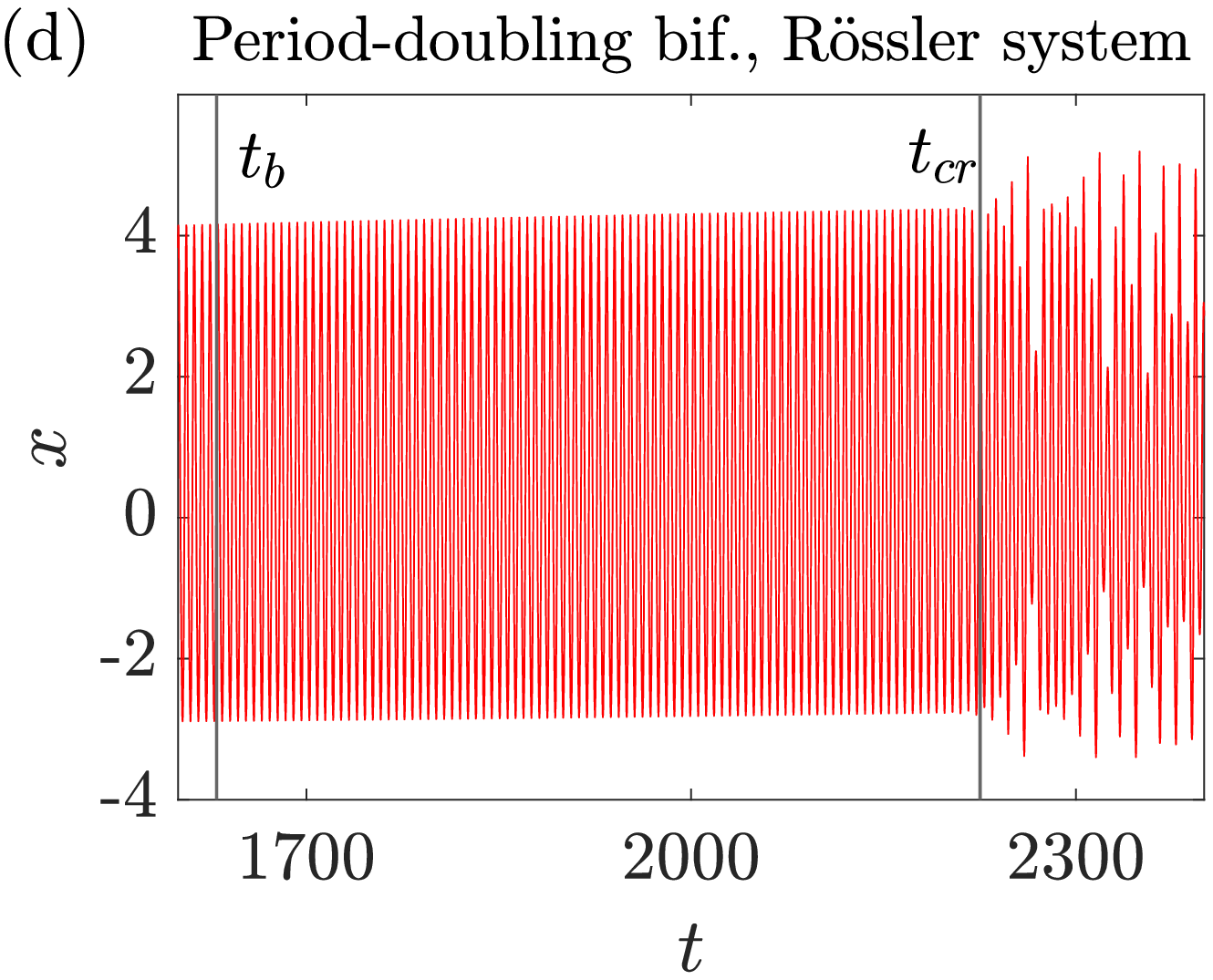}
	\end{center}
	\caption{Time series showing the bifurcation-induced tipping. The parameter drift starts at: (a) $t_{1}=100$, (b) $t_{1}=1000$, (c) $t_{1}=1000$ and (d) $t_{1}=300$. It can be seen that the tipping does not occur at the bifurcation value $t_{b}$, rather it occurs at a posterior value $t_{cr}>t_{b}$. Thus, we observe the shift phenomenon. 
 }
	\label{fig_timeseries}
\end{figure}

Turning to the time-dependent parameter scenario, the time series of a trajectory crossing these bifurcations can be seen in Figs.~\ref{fig_timeseries}(a) and (b). The time at which the parameter $r$ reaches the bifurcation and the critical value ($t_{b}$ and $t_{cr}$) are marked. We consider the condition for tipping to be a threshold of $\eta_{cr}=10^{-2}$ for the distance from the equilibrium (the origin in the case of the pitchfork bifurcation and $C^{\pm}$ in the case of the subcritical Hopf bifurcation). When the threshold is exceeded, the system is considered to change from one regime to another. In both cases, the system tips to another attractor some time after the fixed point looses stability at $t_{b}$.

\subsection{FitzHugh-Nagumo system}

The FitzHugh-Nagumo (FHN) model \cite{FitzHugh1961, Nagumo1962}  is a reduction of the four-dimensional Hodgkin-Huxley model \cite{Hodgkin1952} to a two-variable model. The latter was a very successful model of the initiation and propagation of neural action potential using a squid axon. The simplified version of the FitzHugh-Nagumo model reads
\begin{equation}
	\begin{aligned}
		&\dot{v}=-v(v-a)(v-1)-w+I, \\
		&\dot{w}=b(v-\gamma w),
	\end{aligned}
	\label{FN}
\end{equation}
where the membrane potential $v$ is the main observable, while $w$ models a slow recovery current and $I$ is the magnitude of the stimulus current. The constants $a$, $b$, and $ \gamma $ are kinetic parameters. We fix $ a=0.2, b=0.05, \gamma=0.4 $, and $I$ is going to be the drifting variable, $I(t)$, following Eq.~\eqref{eq_parameter}. 

For the frozen-in version of this system, where $I$ is a fixed parameter, we find a supercritical Hopf bifurcation at $ I_{b} = 0.273 $ \cite{Baer2008}. For $I<I_b$, there is a fixed point $P^{*}=(v^{*},w^{*})$, which satisfies:
\begin{equation}
\begin{aligned}
    &-v^{*}(v^{*}-a)(v^{*}-1)-v^{*}/\gamma+I=0, \\
    &w^{*}=v^{*}/\gamma
\end{aligned}
\label{eq_fixedpointFHN}
\end{equation}
and depends on the value of $I$. A limit cycle appears for $I>I_b$.

If we let $I$ vary with time at small but non-negligible rates and we cross $I_{b}$, we observe the transition from the fixed point given by Eq.~(\ref{eq_fixedpointFHN}) to a periodic solution (see Fig.~\ref{fig_timeseries}(c)). We consider the threshold for the distance from the equilibrium, $P^{*}$, to be $\eta_{cr}=10^{-2}$. Once this threshold is exceeded, the system is considered to have tipped. As can be seen, the tipping occurs at $t_{cr}>t_{b}$. This window of time is the borrowed time, in which the fixed point has lost stability but the system is still tracking it.

\subsection{Rössler system}

The last system we analyze was proposed by Rössler in 1976 \cite{Rossler1976} as a simple model for continuous chaos. The equations read
\begin{equation}
	\begin{aligned}
		&\dot{x}=-y - z, \\
		&\dot{y}=x+ay,\\
		&\dot{z}=b+z(x-c),
	\end{aligned}
	\label{Rossler}
\end{equation}
where $a$, $b$, and $c$ are real parameters. In this case, we take $a$ as the parameter that depends on time, $a(t)$, following Eq.~(\ref{eq_parameter}) and we fix $ b=2 $ and $ c=4 $. 

In the frozen-in version of Eq.~(\ref{Rossler}), where $a$ is fixed, the Rössler system exhibits a transition to chaos through a period-doubling bifurcation, similar to that of the logistic map. For our parameter values, the first transition is found at $ a_{b}=0.333 $.

When we let $a$ vary with time, we observe a transition like the one in Fig.~\ref{fig_timeseries}(d). As can be seen, the periodical trajectory becomes chaotic after $t_{cr}$. The reason for this is that our change rate is too fast to observe all the doublings in between as they are too close. For instance, to observe the next transition at $a=0.374$ the maximum rate is $\varepsilon=10^{-5}$. For later transitions, the rate would decrease. 

In this case, the tipping condition is the loss of periodicity. When the maxima of the time series do not appear periodically, the system is considered to have tipped to the chaotic attractor.

In the following sections, we analyze these systems in both the deterministic and non-deterministic regimes. For the latter, the system approaches the equilibrium exceeding the numerical or experimental precision. Beyond this threshold, noise plays a part and the maximum deviation of the distance from the equilibrium does not vary. This is the case for slow drifts in which the system has time to relax and come closer to the equilibrium. For each regime, we let the parameters evolve with time and we study the shift phenomenon using the described tipping conditions. We describe analytically and numerically the memory effect (dependence of $p_{cr}$ on $p_{0}$) and the rate effect (dependence of $p_{cr}$ on $\varepsilon$).

\section{Deterministic regime} \label{Section_3}

An autonomous dynamical system with constant parameters converges to an attractor after some transient time. However, if the external conditions change, and one of the parameters changes with time, we may observe a stability exchange as the system tips from one attractor to another. 

In this section, we analyze the phenomenon of shifted stability exchange in the deterministic regime, this is, when the precision threshold is not reached and noise does not affect the system's behavior. We analyze the memory and the rate effect in this scenario and we approach this question from two sides: analytical and numerical. 

\subsection{Shifted stability exchange. Analytical approach
\label{sec:ana1}}
We aim to obtain an expression for the shift phenomenon that enable us to calculate the critical value of the parameter $p_{cr}$. Analytically, we deal with the simplest case of a fixed point that looses stability due to the crossing of a bifurcation. We start with the general system
\[
\dot{x}=f(x;p),\quad f(x_{0}(p);p)=0,
\]
where $x_{0}(p)$ is a parameter-dependent equilibrium. The linearization at the equilibrium $x_{0}$ reads
\[
\dot{\xi}=Df(x_{0}(p);p)\xi.
\]

Now, we assume that the system has one leading (critical) eigendirection with
the eigenvector $v(p)$ and the eigenvalue $\lambda(p)$ so that 
\[
Df(x_{0}(p);p)v(p)=\lambda(p)v(p).
\]
 
This assumption holds true close to a pitchfork or transcritical
bifurcation. Thus, from the systems presented above, these calculations are valid only for the pitchfork bifurcation in the Lorenz system. Despite this limitation, the numerical approach will show that the system's response is very similar for more complex bifurcations. We also consider the case that all other eigendirections are stable, and we can restrict the dynamics to the critical direction,
leading to the scalar equation
\begin{equation}
\label{eq:eta}
\dot{\eta}=\lambda(p)\eta,    
\end{equation}
where $\eta$ measures the distance from the equilibrium. 

We remind that the parameter $p$ is a function of time following Eq.~(\ref{eq_parameter}). Thus, during the time interval $t\in [0,t_1]$, the parameter is constant and Eq.~(\ref{eq:eta}) has a constant coefficient $\lambda(p_{0})$. Hence, 
\begin{equation}
\label{eq:eta1}
    \eta(t_{1}) = \eta_0 e^{\lambda(p_{0})t_{1}},
\end{equation}
where $\eta(0)=\eta_0$.
Further, for $t\ge t_{1}$, Eq.~(\ref{eq:eta}) has the form 
\begin{equation}
\label{eq:etaeq}
    	\dot{\eta}=\lambda\left(p_{0}+\varepsilon (t-t_{1})\right) \eta,
\end{equation}
which can be solved as follows: 
\begin{equation}
    \eta(t)=\eta(t_{1}) \exp \left\{ \int_{t_{1}}^{t} \lambda(p_{0}+\varepsilon (t-t_{1})) d t\right\},
\end{equation}
or, equivalently
\begin{equation}
\label{eq:etaeta}
    \eta(t)=\eta(t_{1}) \exp \left\{\frac{1}{\varepsilon} \int_{p_{0}}^{p_{0}+\varepsilon (t-t_{1})} \lambda(p) d p\right\}
    = 
    \eta_0 \exp \left\{
    \lambda(p_0)t_{1} + 
    \frac{1}{\varepsilon} \int_{p_{0}}^{p_{0}+\varepsilon (t-t_{1})} \lambda(p) d p\right\}.
\end{equation}

As previously mentioned, we take the condition for the escape from the equilibrium to be 
\begin{equation}
\label{eq:etacr}
\eta_{cr} = \eta(t_{cr}),    
\end{equation}
where $\eta_{cr}$ is a small threshold, after which the system is considered to be away from the equilibrium, and $t_{cr}$ is the escape time. Using (\ref{eq:etaeta}) and (\ref{eq:etacr}), we obtain 
\begin{equation}
\lambda\left(p_{0}\right) t_1+\frac{1}{\varepsilon} \int_{p_{0}}^{p_{cr}} \lambda(p) d p=\ln \frac{\eta_{cr}}{\eta_{0}}, 
\end{equation}
where $p_{cr} = p_{0}+\varepsilon (t_{cr}-t_{1})$.
Denoting the constant 
\begin{equation}
    \label{eq:C}
    C:=\ln \frac{\eta_{cr}}{\eta_{0}},
\end{equation}
we obtain the following condition for determining the critical parameter value $p_{cr}$:
\begin{equation}
	\int_{p_0}^{p_{cr}} \lambda(p) d p=\varepsilon\left[C-\lambda\left(p_{0}\right) t_{1}\right].
 \label{eq_memory}
\end{equation}

As can be seen, the value of the shift ($p_{cr}-p_{b}$), depends on the initial value of the parameter, $p_{0}$, leading to a memory effect. Furthermore, it depends on the change rate, $\varepsilon$, this is, the rate effect is present as well. 

As an example, we compute Eq.~(\ref{eq_memory}) for the pitchfork bifurcation in the Lorenz system. Now our parameter $p$ is the Rayleigh number, $r$. The eigenvalues for the equilibrium at the origin are
\begin{equation*}
    \begin{aligned}
        &\lambda_{1}=-\beta, \hfill \\
        &\lambda_{2,3}=\frac{1}{2} (-1-\sigma \pm \sqrt{4 r \sigma +\sigma^{2}-2 \sigma+1}).
    \end{aligned}
\end{equation*}

For $r<1 $, all the eigenvalues are negative. At $r=1$, $\lambda_{2}$ changes sign, so the fixed point at the origin becomes unstable giving rise to the pitchfork bifurcation.  Thus, $r_{b}=1$ and the critical eigenvalue in this case is $\lambda_{2} =: \lambda$. For our parameter values, it reads
\begin{equation}
    \lambda=\frac{1}{2}\left(-11+\sqrt{81+40r} \right).
    \label{eigenvalue}
\end{equation}

Now, integrating the eigenvalue as in Eq.~(\ref{eq_memory}) we get:
\begin{equation}
\begin{aligned}
& \frac{1}{2} \int_{r_0}^{r_{cr}}(-11+\sqrt{81+40 r})dr=\\
& =\frac{-11\left(r_{cr}-r_0\right)}{2}+\frac{1}{120} \cdot\left[\left(40 r_{cr}+81\right)^{3 / 2}-\left(40 r_0+81\right)^{3 / 2}\right].
\end{aligned}
\end{equation}

As we are interested in representing $r_{1}=r_{cr}-r_{b}$ against $r_{2}=r_{b}-r_{0}$, this is, the distances before and after the bifurcation, we apply a change of variables so that $r_{cr}=r_{1}+r_{b}$ and $r_{0}=r_{b}-r_{2}$. Finally, using Eq.~(\ref{eq_memory}) we obtain:
\begin{equation}
\begin{aligned}
  & \frac{-11\left(r_{1}+r_{2}\right)}{2}+\frac{1}{120} \cdot\left[\left(40 (r_{1}+r_{b})+81\right)^{3 / 2}-\left(40 (r_{b}-r_{2})+81\right)^{3 / 2}\right] =\\
  & =\varepsilon \left[ C-\frac{-11 \sqrt{81+40 (r_{b}-r_{2})}}{2}  t_1\right].
  \end{aligned}
  \label{eq_memoryLorenz}
\end{equation}

\begin{figure}[h]
	\begin{center}
		\includegraphics[width=0.45\textwidth]{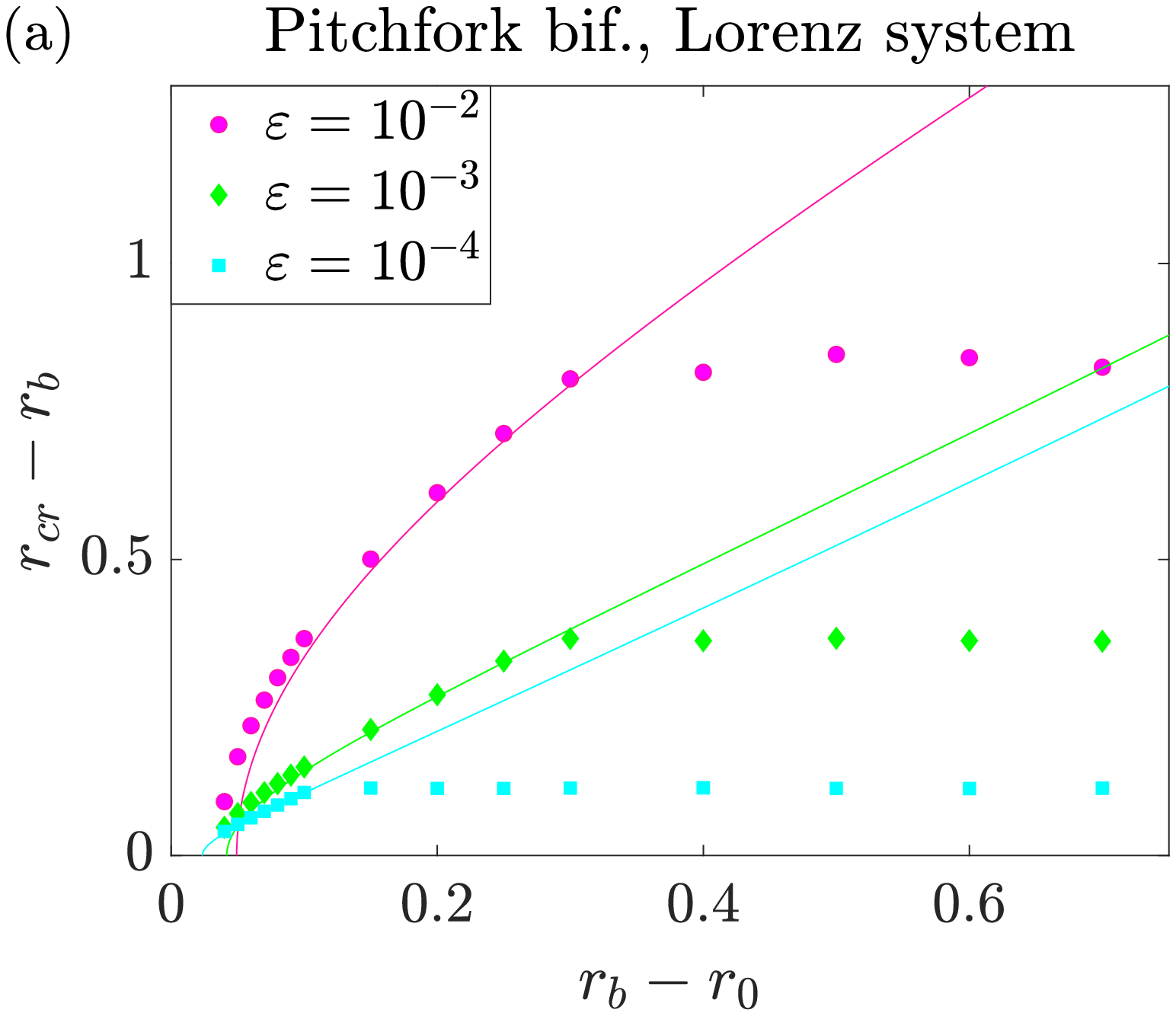}
		\includegraphics[width=0.45\textwidth]{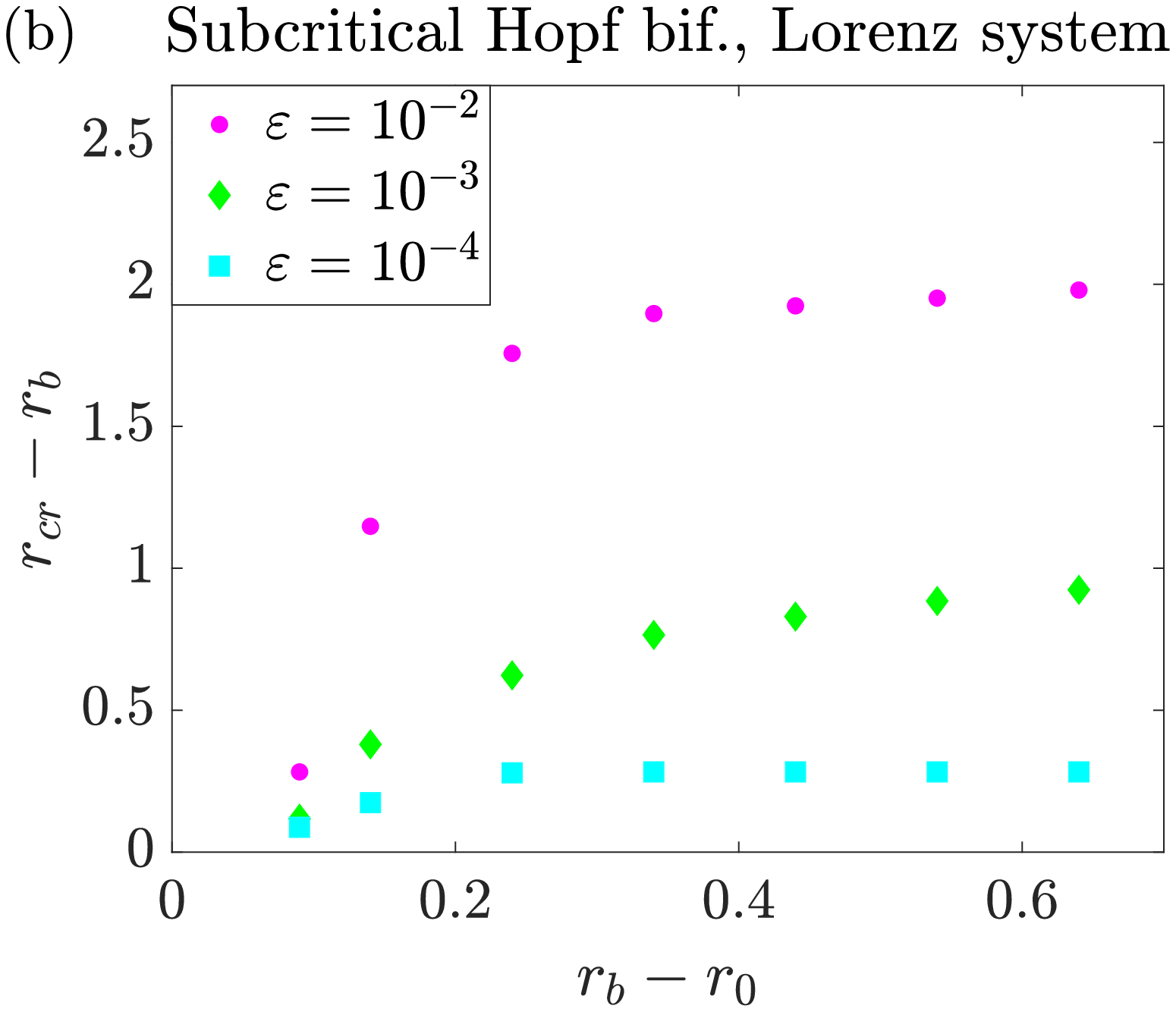}
		\includegraphics[width=0.45\textwidth]{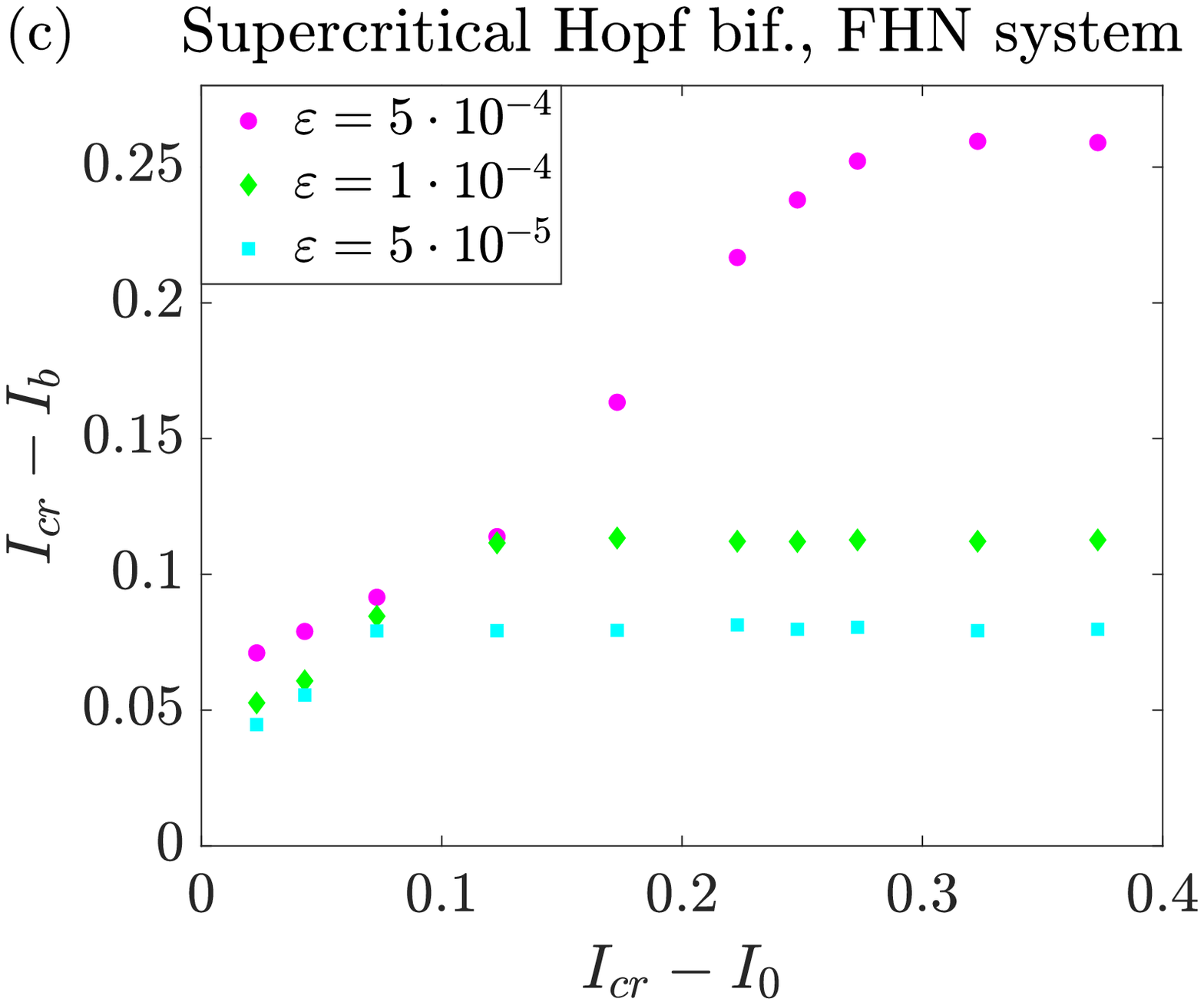}
		\includegraphics[width=0.45\textwidth]{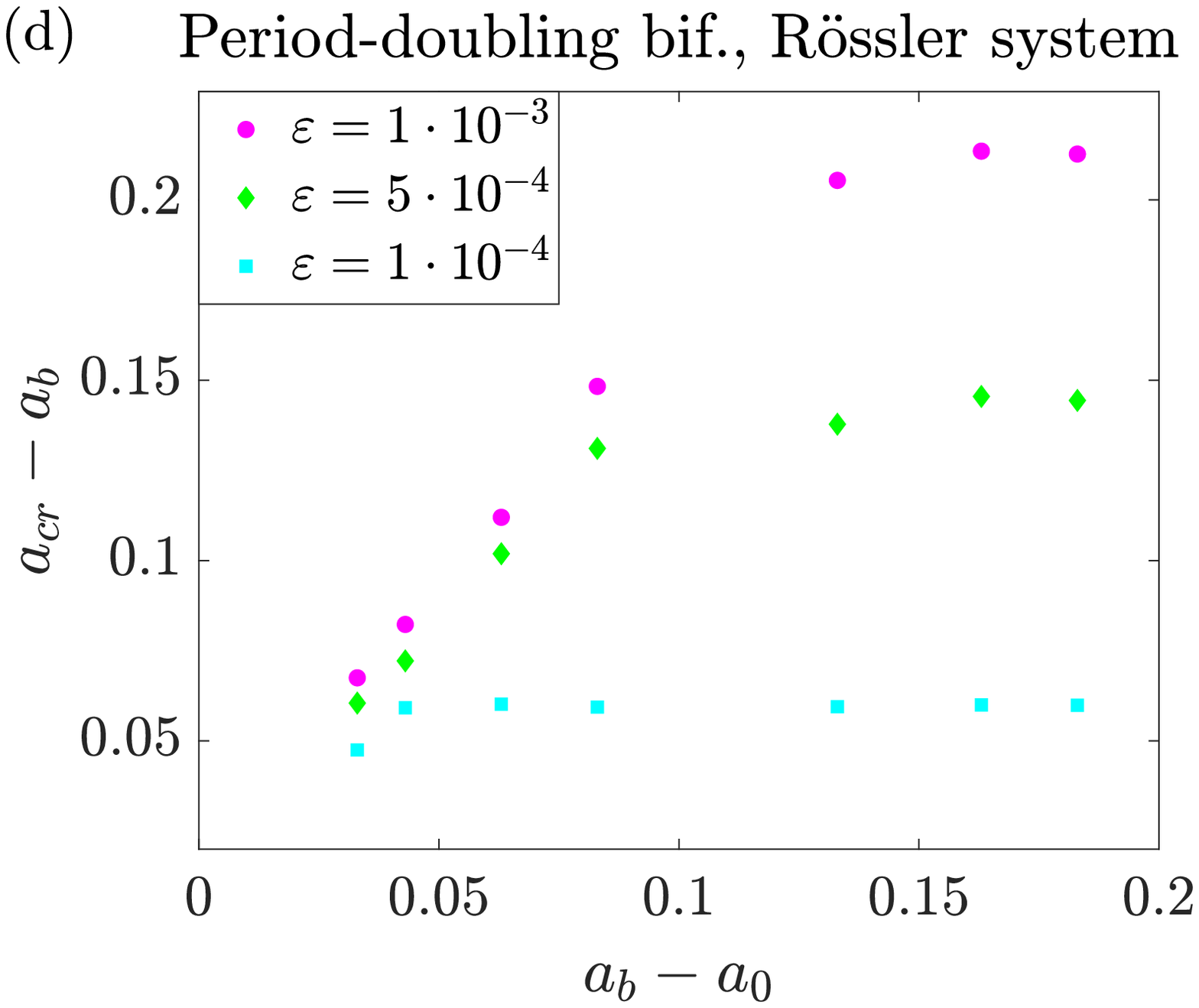}
	\end{center}
	\caption{Memory and rate effects for different bifurcations and systems in the deterministic regime. The curves in (a) are analytically calculated, while the points in (a), (b), (c) and (d) are numerically calculated. The difference in both approaches comes from the impact of the numerical noise that causes the memory loss as we enter the non-deterministic regime. This can be seen in the form of plateaus. Before that, in the deterministic regime, both approaches can be seen to match for the pitchfork bifurcation. }
	\label{fig_memory}
\end{figure}

To obtain the value of $C$, we take the initial distance to the equilibrium $\eta_{0}=1$ and the critical distance $\eta_{cr}=10^{-2}$, following the criterion presented in Section \ref{Section_2}. We also fix the time at which the parameter starts drifting to $t_{1}=100$. 
Equation~\eqref{eq_memoryLorenz} can be numerically solved to obtain the curves in Fig.~\ref{fig_memory}(a) for different values of $\varepsilon$.

The memory effect is clear in the figure: for initial values of the parameter closer to the bifurcation value, the tipping occurs before. This relation is not linear in general. 
It is linear only for the case of simple systems in which the eigenvalue $\lambda$ is linearly dependent on the parameter: $\lambda(p)=m p$, where $m \in \mathbb R$. 
This is not the case of the Lorenz system, see Eq.~\eqref{eigenvalue}. The linearization of the eigenvalue was considered in \cite{Baer2008} taking a small value of $\varepsilon$, but here we obtained the general expression.

The rate effect can also be seen in Fig.~\ref{fig_memory}(a): for a fixed value of $r_{0}$, the shift increases with the change rate. The shift phenomenon has been considered to be independent of $\varepsilon$ in some previous works as for initial values of the parameter very close to the bifurcation, the curves come closer. However, we show that this does not hold true when we start away from the bifurcation and when we allow the system to have an initial transient time $t_1$ before the drift starts. 

\subsection{Shifted stability exchange. Numerical approach}

Now, we analyze the shifted stability exchange numerically. However, we restrain to the case that the numerical threshold is not reached and the dynamics is deterministic. We integrate Eqs.~\eqref{Lorenz}, \eqref{FN}, and \eqref{Rossler} with the parameter drift starting at  (a) $t_{1}=100$, (b) $t_{1}=1000$, (c) $t_{1}=1000$, and (d) $t_{1}=300$, for different rates $\varepsilon$. These values are chosen to skip the transient behavior before the drift starts. We show the shift phenomenon for all the bifurcations in Fig.~\ref{fig_memory}. The values of $\varepsilon$ and $p_{0}$ are also different from one system to another, attending to the natural time scale and the presence of other bifurcations that limit the maximum value of $p_{cr}-p_{0}$.

Comparing the analytical and numerical data for the pitchfork bifurcation, it can be seen that both approaches match for a certain region, but afterwards the numerical data yields a plateau. This plateau, present in all the other bifurcations too, corresponds to the non-deterministic regime. In that regime we let the system time enough to relax and approach the attractor beyond the numerical precision. This can be the case for slow enough rates or values of $r_{0}$ far enough from the bifurcation. In these cases, the numerical noise plays the role of a memory-loss agent. We explore the non-deterministic regime in more detail in the following section. 

Notice that in the deterministic regime all the systems' responses are quite similar to what was predicted by Eq.~(\ref{eq_memory}). The analytical calculation is valid for a system that has one leading (critical) eigendirection, which holds true for such bifurcations as the pitchfork. However, we can see that even for more complex bifurcations as the period-doubling, we obtain a similar memory effect. In all cases, the shift increases with $\varepsilon$ and also increases with the initial distance to the bifurcation before entering the non-deterministic regime.

\begin{figure}[h]
	\begin{center}
		\includegraphics[width=0.45\textwidth]{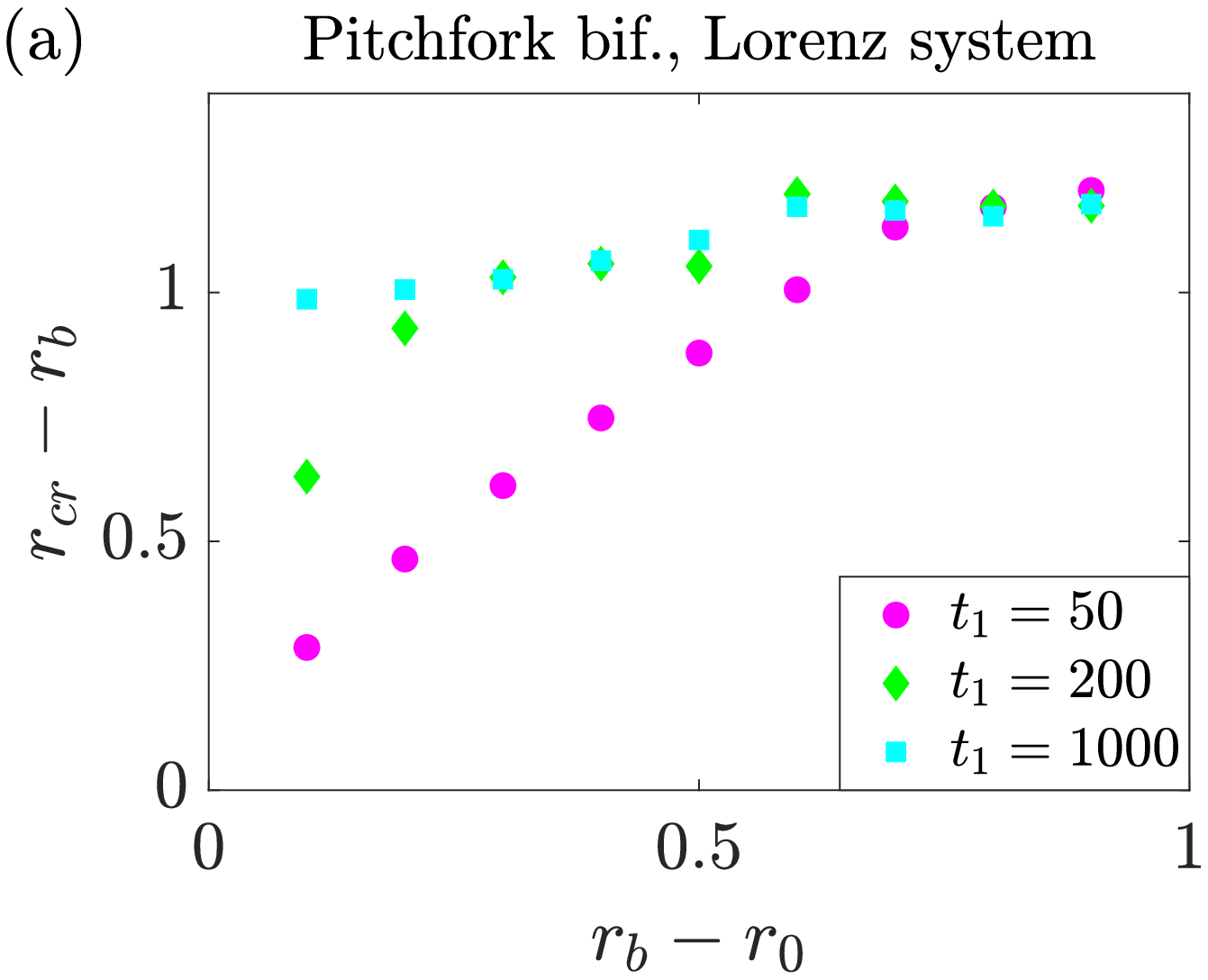}
		\includegraphics[width=0.45\textwidth]{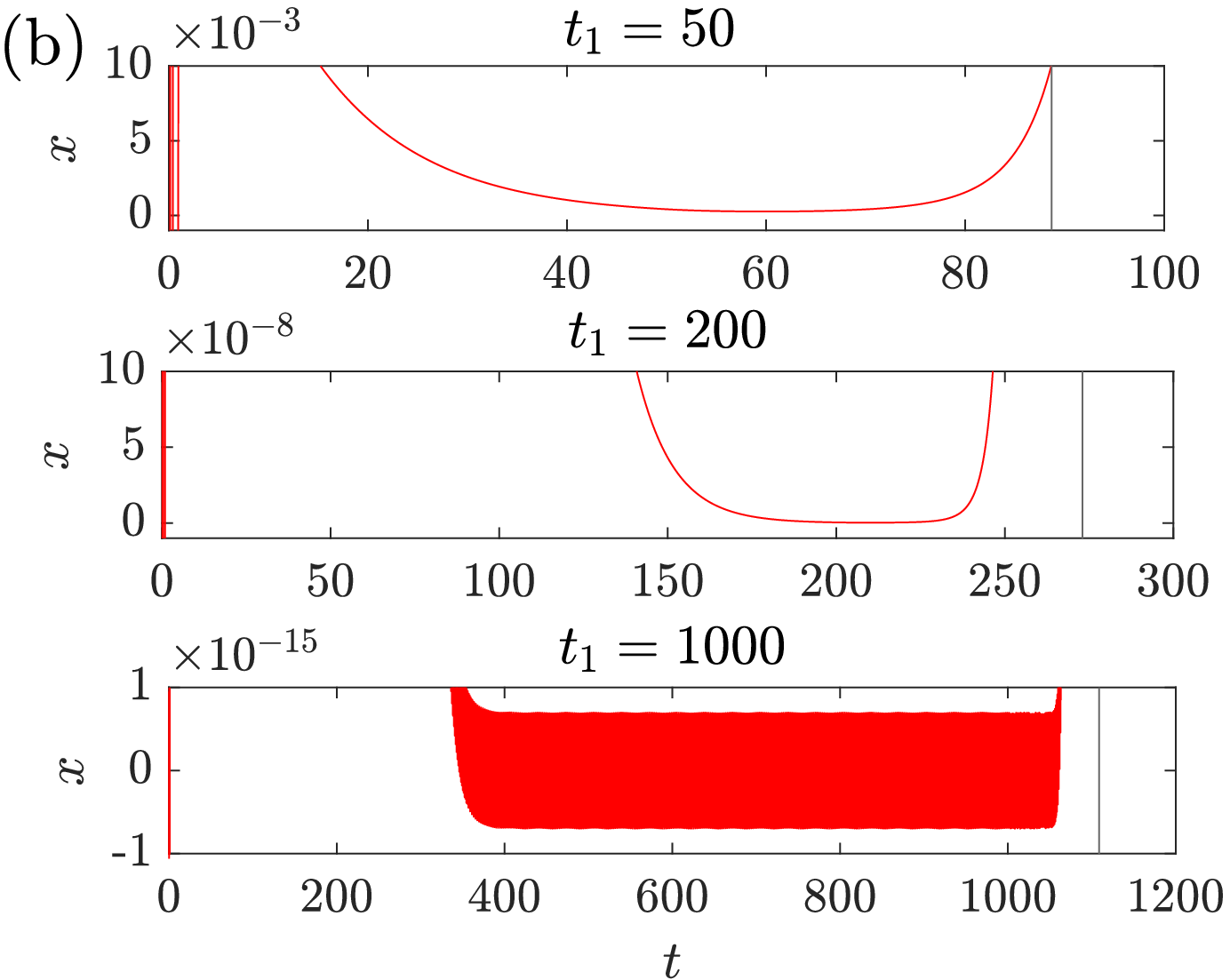}
	\end{center}
	\caption{Memory effect for different values of $t_{1}$. In (a) it can be seen that the shift is less pronounced for small values of $t_{1}$. For values of $r_{0}$ further enough from the bifurcation, this effect disappears ($r_{cr}$ becomes independent of $t_{1}$) as the three trajectories enter the non-deterministic region. In (b) we show the time series for $r_{0}=0.9$, this is, $r_{b}-r_{0}=0.1$. Notice that for $t_{1}=1000$, numerical noise is already present and causes the memory loss. }
	\label{memory_t1}
\end{figure}

Another parameter that may seemingly affect the shift phenomenon is $ t_{1} $. It plays a role similar to $ \varepsilon $. For small values of $ t_{1} $, the system has no time to relax before the parameter drift starts, while for large $ t_{1} $, we enter again the non-deterministic regime and we observe a memory loss. This can be seen in Fig.~\ref{memory_t1} for the pitchfork bifurcation and $\varepsilon=10^{-2}$. On the left panel, we observe how the memory loss appears even for small values of $ t_{1} $ when the initial value of the parameter is far enough from the bifurcation, allowing the system to relax once again. On the right side, we can see how the different trajectories approach the fixed point at the origin for $r_{0}=0.9$; this is, $r_{b}-r_{0}=0.1$. Notice the different orders of magnitude in the $ y- $axis and the presence of numerical noise for the largest value of $t_{1}$. For $t_{1}=50, 200$, the trajectory is deterministic, while for $t_{1}=1000$, we enter the non-deterministic regime. 

For the other bifurcations, changing $ t_{1} $ barely affects the shift as the value of the equilibrium also depends on the value of the parameter, see Eq.~\eqref{eq_fixedpoint_Lorenz} or \eqref{eq_fixedpointFHN}. Note that, on the contrary, for the pitchfork bifurcation the equilibrium is fixed at the origin for $r<r_{b}$. This implies that for the other bifurcations it is more difficult to track the equilibrium for $t_{1}<t<t_{b}$. Thus, in those cases getting closer to the equilibrium before the parameter drift starts is not enough to observe the memory loss. 
 
\section{Non-deterministic regime} \label{Section_4}

In this section we explore the shifted stability exchange when the system enters the non-deterministic regime. This means that the system reaches the precision threshold and noise affects the dynamics causing a memory loss. We present analytical and numerical results for the systems presented in Section \ref{Section_2} as well. 

\subsection{Shifted stability exchange. Analytical approach} \label{sec:non-det-A}

We aim to obtain an analytical expression for $p_{cr}$ that takes into account that the system reaches a threshold  $\eta_n=\eta(t_n)$ of the numerical precision within the observable time interval $t_n<t_{cr}$. Depending on whether the threshold is reached before or after the drift starts, we can classify the system's behavior in two cases, A and B. 

    \begin{itemize}
        \item[(A)]  The system reaches the numerical precision threshold before the parameter drift starts ($t_1>t_n$). In this case, the dynamics of the system can be divided into the following steps:
        \begin{itemize}
            \item[(i)]  Exponential convergence to the equilibrium with  fixed rate $\lambda(p_{0})$ until $t_{n}$.
            \item[(ii)]  Drifting phase within the numerical precision error for $t\in [t_n,t_{b}]$, $t_{b}>t_1$. 
            \item[(iii)]  Non-autonomous deterministic dynamics according to Eq.~\eqref{eq:etaeq} on $[t_{b},t_{cr}]$, until the escape threshold $\eta_{cr}=\eta(t_{cr})$  is reached.
        \end{itemize}
        \item[(B)] The threshold is not reached before the parameter drift starts ($t_1<t_n$). In this case, the dynamics is divided into the following steps:
        \begin{itemize}
            \item[(i)]  Exponential convergence to the equilibrium with  fixed rate $\lambda(p_{0})$ until $t_{1}$.
            \item[(ii)] 
            Non-autonomous deterministic dynamics according to Eq.~\eqref{eq:etaeq} on $[t_{1},t_n]$, until the numerical precision threshold $\eta_n=\eta(t_n)$  is reached.
            \item[(iii)]  Drifting phase within the numerical precision error for $t\in [t_n,t_{b}]$, $t_{b}>t_1$.             
            \item[(iv)]  Non-autonomous deterministic dynamics according to Eq.~\eqref{eq:etaeq} on $[t_{b},t_{cr}]$, until the escape threshold $\eta_{cr}=\eta(t_{cr})$  is reached.
        \end{itemize}
    \end{itemize}

In both cases, the system starts feeling the repulsion at the bifurcation value, this is, when the equilibrium looses stability. From this moment, the dynamics becomes deterministic again. 

Starting from the reduced equation for the distance from the equilibrium, Eq.~\eqref{eq:etaeq}, and taking into account Eqs.~\eqref{eq:eta1}--\eqref{eq:etaeta}, we obtain the following expressions for the distance $\eta(t)$ for case A: \\
(A-i):
\begin{equation}
    \label{eq:Ia}
    \eta_n = \eta_{0} e^{\lambda(p_{0})t_{n}},
\end{equation}
(A-ii):
\begin{equation}
    \label{eq:IIa}
    \eta(t_{b}) = \eta_b= \eta_n= \eta_0 e^{\lambda(p_{0})t_{n}},
\end{equation}
(A-iii):
\begin{equation}
    \label{eq:IIIa}
        \eta(t)=\eta(t_{b}) \exp \left\{ \int_{t_{b}}^{t} \lambda(p_{0}+\varepsilon (t-t_{1})) d t\right\}.
\end{equation}

During phase (A-ii), the maximum deviation does not vary, and there are fast fluctuations related to the numerical noise and particularities of the numerical integration scheme.  We neglect these fluctuations and only take into account that the distance starts to grow exponentially after this phase from $\eta_{b}$.  Equivalently, the integral in Eq.~(\ref{eq:IIIa}) can be written in terms of the parameter instead of time, leading to
\begin{equation}
    \label{eq:IIIa-1}
        \eta(t)=\eta(t_{b}) \exp \left\{\frac{1}{\varepsilon} \int_{p_{0}+\varepsilon(t_{b}-t_1)}^{p_{0}+\varepsilon (t-t_{1})} \lambda(p) d p\right\}
    = 
    \eta_0 \exp \left\{
    \lambda(p_{0})t_{n} + 
    \frac{1}{\varepsilon} \int_{p_{0}+\varepsilon(t_{b}-t_{1})}^{p_{0}+\varepsilon (t-t_{1})} \lambda(p) d p\right\}.
\end{equation}

Now, for $t=t_{cr}$, we apply the escape condition 
\begin{equation}
    \ln \frac{\eta_{cr}}{\eta_{0}} = 
    \lambda(p_{0})t_n + 
    \frac{1}{\varepsilon} 
    \int_{p_{b}}^{p_{cr}} 
    \lambda(p) d p
\end{equation}
with $p_{cr} = p_0+\varepsilon (t_{cr}-t_1)$ and $p_{b} = p_0+\varepsilon(t_{b}-t_1)\ge p_0$. Using Eq.~(\ref{eq:Ia}), we denote $\lambda(p_0)t_n = \ln \frac{\eta_n}{\eta_0}:= C_1$. Finally, taking into account Eq.~(\ref{eq:C}), we get 
\begin{equation}
    \label{eq:IIIa-escape}
    \int_{p_{b}}^{p_{cr}} 
    \lambda(p) d p  = 
    \varepsilon [C-C_1].
\end{equation}

From this expression we can obtain the value of $p_{cr}$, which  does not depend on $p_{0}$. Thus, we observe a memory loss due to the presence of noise.  This was already observed in Fig.~\ref{fig_memory} for the numerically calculated points: when the system entered the non-deterministic regime the points reached a plateau indicating that the exchange of stability is independent of $p_{0}$. On the other side, Eq.~(\ref{eq:IIIa-escape}) does depend on the change rate. Integrating this equation for the pitchfork bifurcation in the Lorenz system and considering $\eta_{n}=10^{-19}$, we depict this dependence in Fig.~\ref{scalelaw}(a) as a red curve. It is represented in logarithmic scale for a better visualization. This corresponds to representing the different heights of the plateaus from Fig.~\ref{fig_memory}.  We observe that the shift increases with increasing change rate.

Another clear implication of Eq.~\eqref{eq:IIIa-escape} is that $p_{cr}\to p_b$ as $\varepsilon\to 0$. This implies that as the shift tends to zero, we recover the frozen-in bifurcation diagram. This would be the case of a quasi-static parameter variation. Furthermore, for sufficiently small $\varepsilon$, a linearization $\lambda(p) \approx \lambda_0' (p-p_b)$ can be used, leading to 
$$
p_{cr}=p_{b}+\sqrt{\frac{2\varepsilon}{\lambda'_{0}}\left[C-C_{1}\right]},
$$
which implies the square root scaling $(p_{cr}-p_b) \sim \varepsilon^{1/2}$. This power law has been drawn in Fig.~\ref{scalelaw} as dashed blue lines for all the bifurcations. For the pitchfork bifurcation it is very similar to the read curve calculated using Eq.~\eqref{eq:IIIa-escape}. And for the rest of the bifurcations, the numerically calculated points seem to match this approximation too. All in all, we find similar qualitative behavior for different bifurcations and systems following the power law $(p_{cr}-p_b) \sim \varepsilon^{1/2}$.

For case B, we obtain analogously the following\\
(B-i):
\begin{equation}
    \label{eq:Ib}
    \eta_1 = \eta_0 e^{\lambda(p_0)t_1},
\end{equation}
(B-ii):
\begin{equation}
    \label{eq:IIb}
        \eta_n = 
        \eta(t_n)=
        \eta_1 \exp 
        \left\{ \int_{t_1}^{t_n} 
        \lambda(p_0+\varepsilon (t-t_1)) 
        d t
        \right\},
\end{equation}
\begin{equation}
\label{eq:etaeta2}
    \eta_n=
    \eta_0 \exp \left\{
    \lambda(p_0)t_1 + 
    \frac{1}{\varepsilon} 
    \int_{p_0}^{p_n} \lambda(p) d p
    \right\},
\end{equation}
where $p_n = p_0 +\varepsilon (t_n-t_1)$,\\
(B-iii): 
\begin{equation}
    \label{eq:IIIb}
    \eta_{b} = \eta(t_{b}) = \eta_n,
\end{equation}
and (B-iv):
\begin{equation}
    \label{eq:IVb}
    \eta_{cr} = \eta_{b} 
    \exp \left\{
    \frac{1}{\varepsilon} 
    \int_{p_{b}}^{p_{cr}} 
    \lambda(p) dp
    \right\} 
    = 
    \eta_0
    \exp \left\{
    \lambda(p_0)t_1 + 
    \frac{1}{\varepsilon} 
    \left(
    \int_{p_0}^{p_n} \lambda(p) d p + 
    \int_{p_{b}}^{p_{cr}}  \lambda(p) dp
    \right)
    \right\},    
\end{equation}
leading to 
\begin{equation}
    \label{eq:IVb-escape}
    \int_{p_0}^{p_n} \lambda(p) d p + 
    \int_{p_{b}}^{p_{cr}}  \lambda(p) dp
    =
    \varepsilon
    \left[
    C
    -
    \lambda(p_0)t_1 
    \right].
\end{equation}
In this last equation, the integrals correspond to the deterministic part of the drift before reaching the numerical threshold and after the bifurcation is reached. The difference of Eq.~(\ref{eq:IVb-escape}) from Eq.~(\ref{eq_memory}) is that the interval $[p_{n},p_{b}]$ must not be taken into account due to the presence of numerical noise. In this interval, the maximum deviation from the equilibrium stays the same. In case B, we find again the memory effect as Eq.~(\ref{eq:IVb-escape}) depends on $p_{0}$.

\subsection{Shifted stability exchange. Numerical approach}

Now, we numerically calculate $p_{b}-p_{cr}$ for the non-deterministic regime. The values of $t_{1}$ remain the same as in the previous section and we fix $p_{0}$ to the following values: (a) $r_{0}=0.5$, (b) $r_{0}=24.5$, (c) $I_{0}=0.05$, (d) $a_{0}=0.2$, that belong to the beginning of the plateaus in Fig.~\ref{fig_memory}, thus we make sure that we are in the non-deterministic regime. For these values, in all cases we find that $t_{1}<t_{n}$, which is case A from Section~\ref{sec:non-det-A}. 

\begin{figure}[h]
	\begin{center}
		\includegraphics[width=0.45\textwidth]{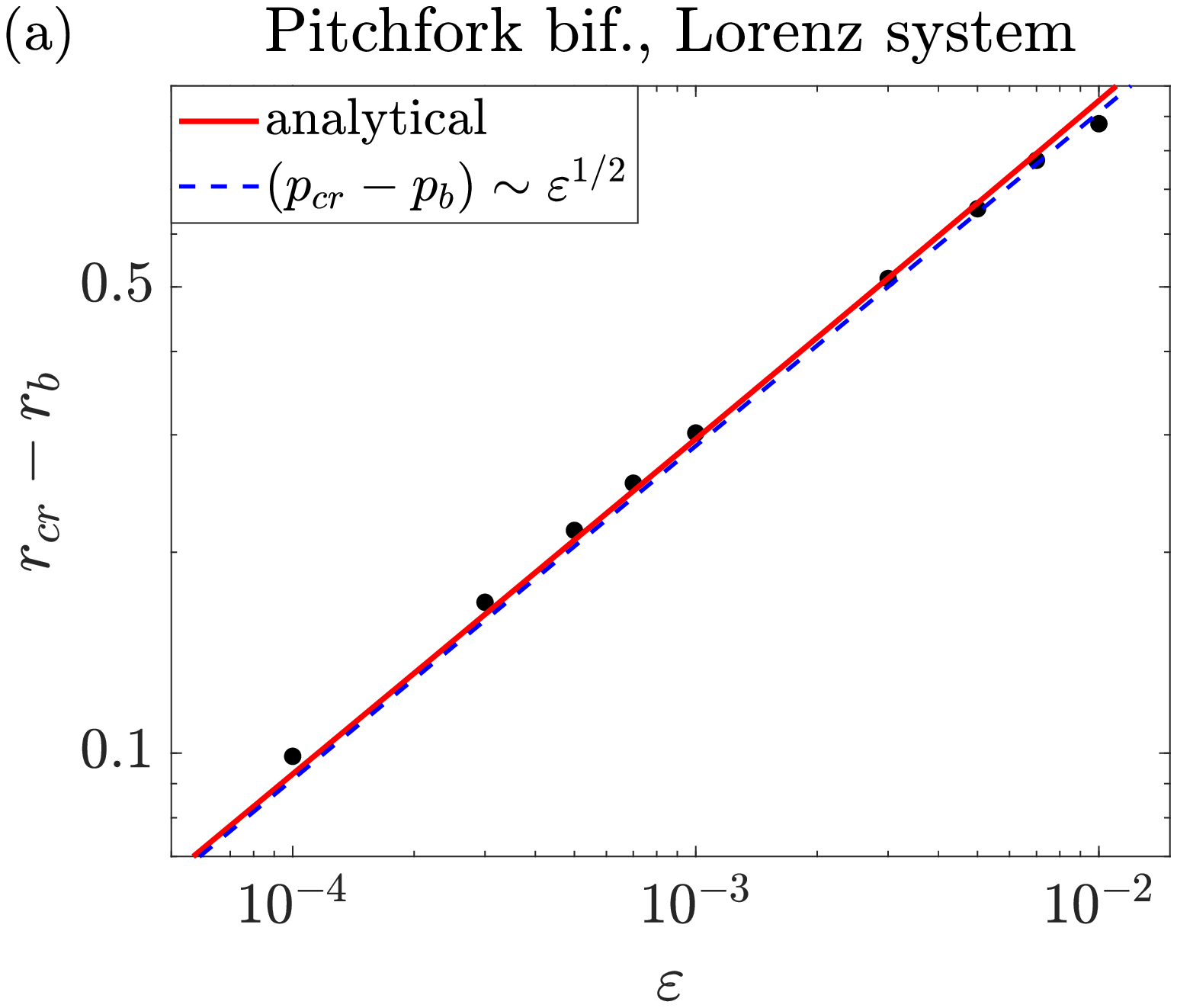} 
		\includegraphics[width=0.45\textwidth]{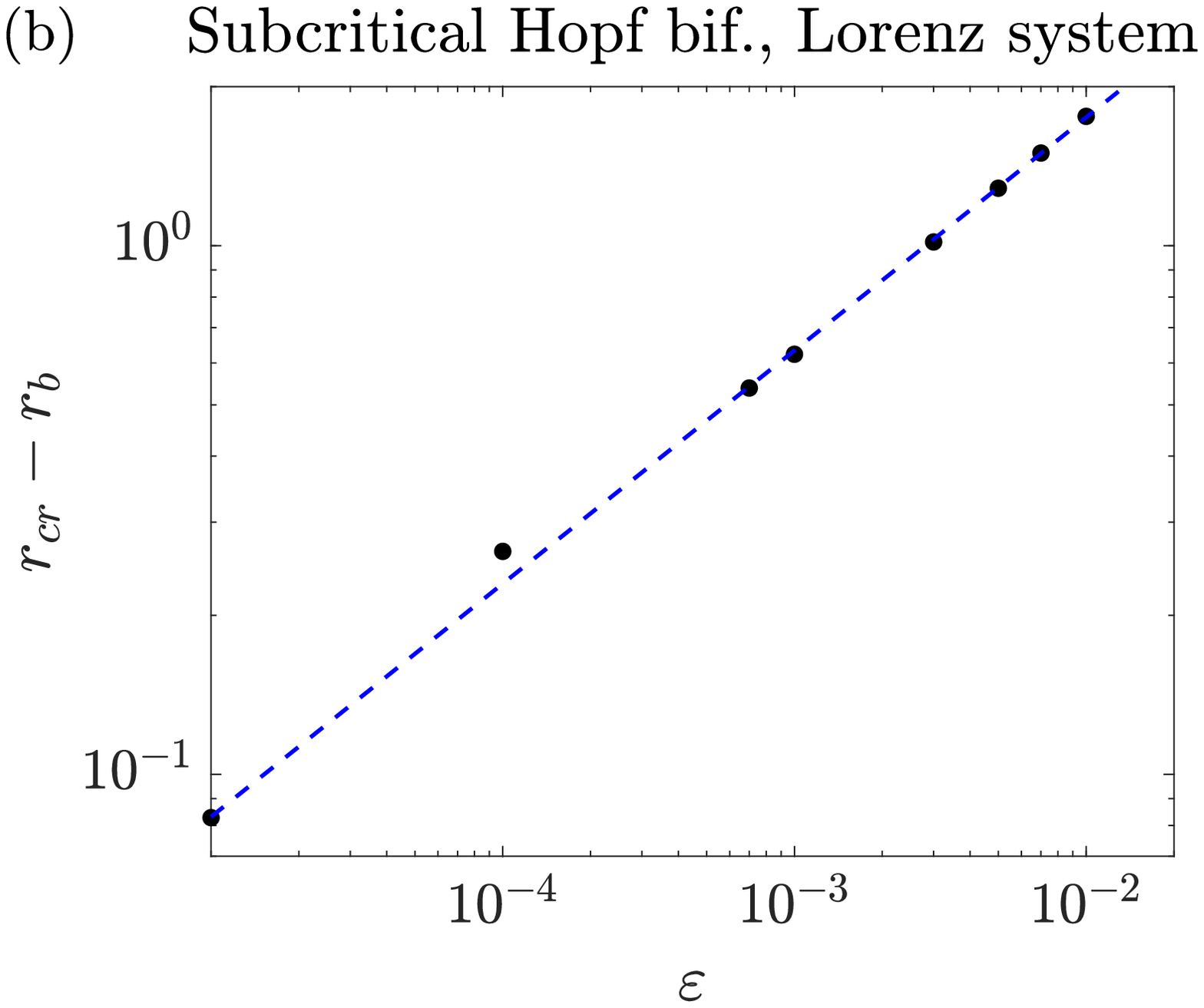} \\
        \includegraphics[width=0.45\textwidth]{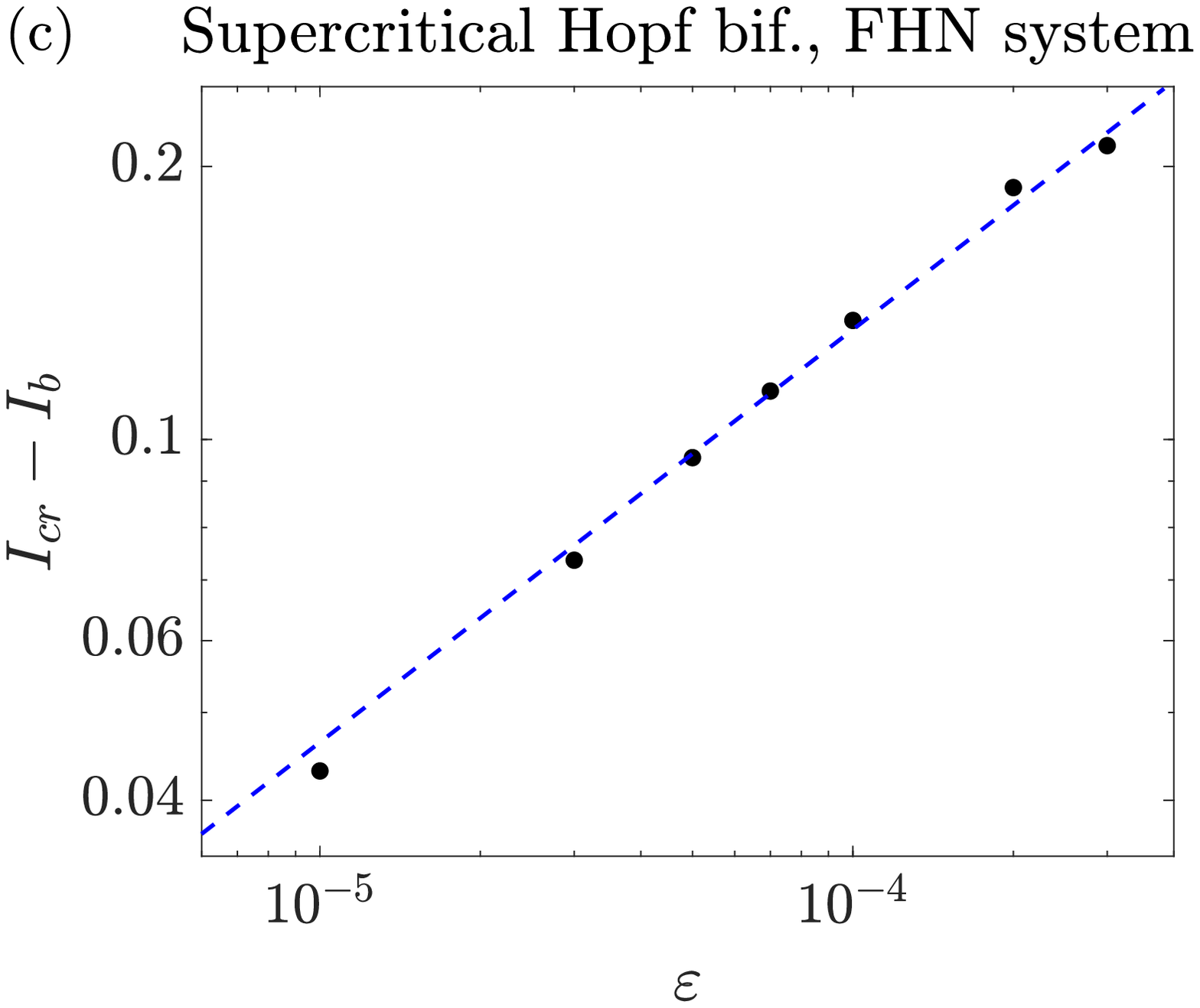}
		\includegraphics[width=0.45\textwidth]{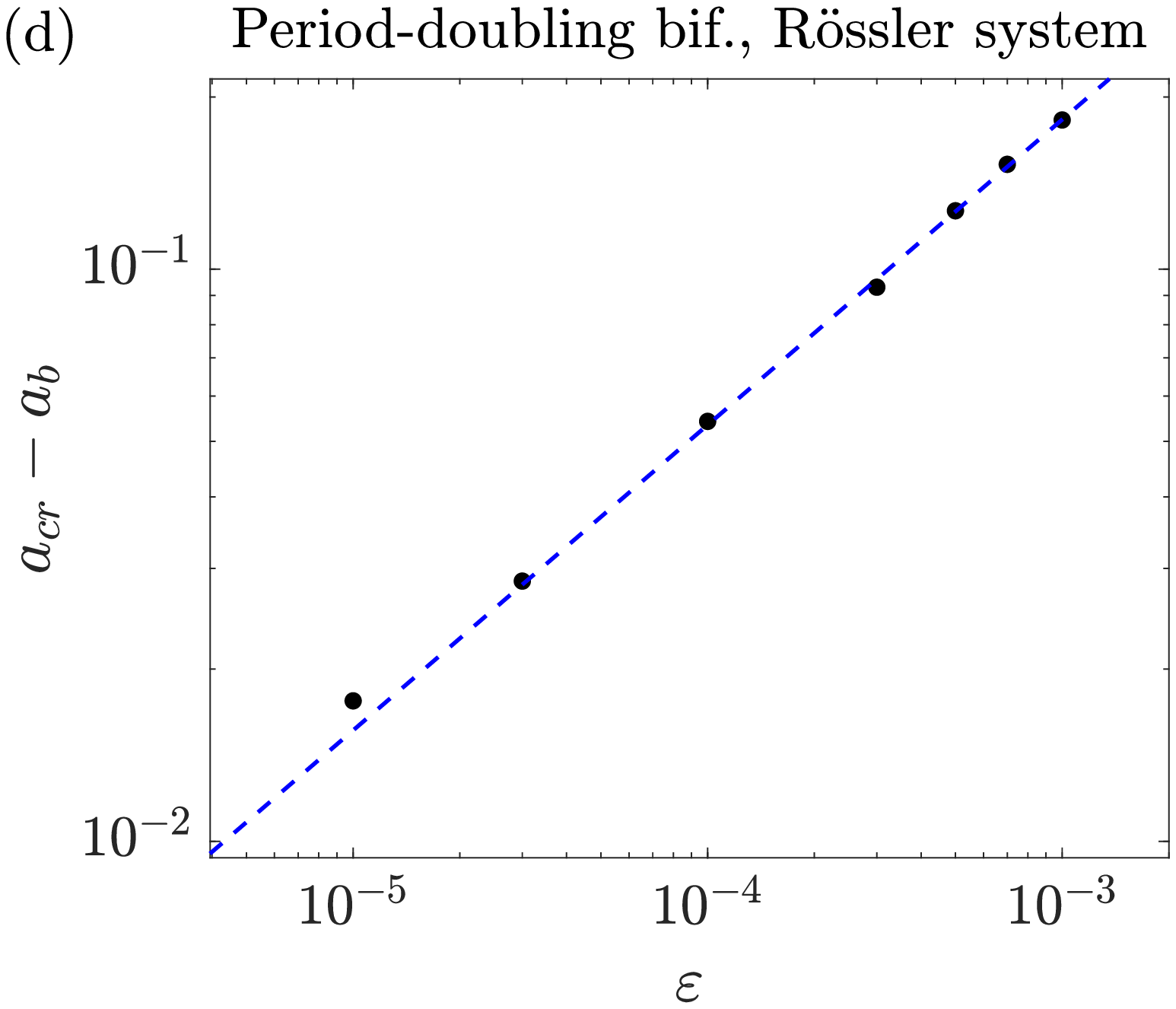}
	\end{center}
	\caption{Rate effect on the shifted stability exchange for different bifurcations and systems. The plots are in logarithmic scale for a better visualization. The red curve in (a) is analytically calculated from Eq.~(\ref{eq:IIIa-escape}) and the points in (a), (b), (c), and (d) are numerically calculated. The response is similar in all systems: for higher change rates, the shift is larger and it seems to follow approximately a power law $(p_{cr}-p_b) \sim \varepsilon^{1/2}$ in all cases (dashed blue lines).}
	\label{scalelaw}
\end{figure}

As previously mentioned, we find a memory loss in this regime, which can be seen in the form of plateaus in Fig.~\ref{fig_memory}. Regarding the rate effect, we present the results in Fig.~\ref{scalelaw}. The points are depicted in logarithmic scale for a better visualization. As stated in the previous section, these numerically calculated points match the power law $(p_{cr}-p_b) \sim \varepsilon^{1/2}$ (dashed blue lines). This implies that no matter the bifurcation or the system, we observe a similar response to a variation in the change rate.

These results are in agreement with previous work dealing with complex bifurcations such as the heteroclinic bifurcation in the Lorenz system \cite{Cantisan2021} and the boundary crisis in the Duffing oscillator with delay \cite{Cantisan2021a}. In both cases, chaotic attractors that loose stability are involved, thus they are not expected to follow our analytical calculations (the exponent in the power law differs from $1/2$). However, qualitatively they present similar rate effects.

In Table \ref{table} we summarise the different effects depending on the regime in which the system is found. For the deterministic regime and non-deterministic regime with $t_{1}<t_{n}$, the shift depends on the initial value of the parameter $p_{0}$ and on $\varepsilon$. In particular, we observe that as $|p_{b}-p_{0}| \rightarrow 0$, the dependence on $\varepsilon$ is less pronounced. On the other hand, for the non-deterministic regime with $t_{1}>t_{n}$  memory loss occurs. Thus, the shift is independent of $p_{0}$ and we can approximate the shift as $(p_{cr}-p_b) \sim \varepsilon^{1/2}$.

\begin{table}
\begin{tabular}{|>{\centering}p{5cm}|>{\centering}p{5cm}|>{\centering}p{5cm}|}
\hline 
\multirow{2}{5cm}{\textbf{Determininistic regime}} & \multicolumn{2}{>{\centering}p{10cm}|}{\textbf{Non-determininistic regime}}\tabularnewline
\cline{2-3} \cline{3-3} 
 & $t_{1}>t_{n}$ & $t_{1}<t_{n}$\tabularnewline
\hline 
Memory effect 
 & No memory effect 
 & Memory effect 

\tabularnewline
\hline 
Rate effect 

 & Rate effect \quad \quad
 $(p_{cr}-p_b) \sim \varepsilon^{1/2}$ & Rate effect 

\tabularnewline
\hline 
\end{tabular}
 \caption{Summary of the memory ($p_{cr}$ dependence on $p_{0}$) and rate ($p_{cr}$ dependence on $\varepsilon$) effects for the deterministic and non-deterministic regimes.}
 \label{table}
\end{table}

%\FloatBarrier
\section{Conclusions and discussion} \label{Section_5}

We have studied the shifted stability exchange that occurs when a drifting parameter crosses a bifurcation point. This phenomenon causes that, the tipping from one attractor to another appears for a parameter value beyond the bifurcation point. To this end, we have analyzed some paradigmatic systems to find common features of this phenomenon. We found that there are two regimes: deterministic and non-deterministic. For certain drift scenarios, when the rate of change of the parameter is sufficiently low, the system comes very close to an attractor, so that the precision (numerical noise) does not allow it to be attracted any further. In this case, noise plays an important role and leads to the non-deterministic regime. This is a very common scenario as all numerical algorithms are subject to precision limits. Furthermore, this is also the regime of an experimental setup that has precision limits when approaching the attractor. 

For the deterministic regime, we have derived the expression for the shift phenomenon, i.e., the value of $p_{cr}$, which depends on the initial value of the parameter (memory effect) and on the change rate (rate effect). In this case, the rate effect is more intense for initial values of the parameter further from the bifurcation. This expression is valid for bifurcations such as the pitchfork bifurcation, but the numerical calculations show similar qualitative behavior for more complex bifurcations such as the period-doubling bifurcation.

For the non-deterministic regime, we have distinguished between the case where the numerical threshold is reached before and after the drifting. When it is reached before, the noise acts as a memory-loss agent. We have derived the analytical expression for the shift and proved that it matches the numerical calculations too. For the rate effect we observe a power law scaling of the type: $(p_{cr}-p_b) \sim \varepsilon^{1/2}$ for all the systems and bifurcations. This implies that if $\varepsilon\to 0$, then $p_{cr}\to p_b$, thus recovering the frozen-in bifurcation diagram. For the case that the threshold is reached after the drifting, we can see a similar response to the one in the deterministic regime. We have also derived the analytical expression for this case.

Finally, we address the implications of our work. There is a lot of research that aims to predict when a system is going to tip, ending the so-called borrowed time between the bifurcation and the tipping. Sometimes this transition is undesirable as it might be the case of population extinction in the context of ecology or climate change due to anthropogenic causes. In other situations, the tipping might be desirable, but the shift should be minimized as much as possible. This is the case of some physical experiments, such as transitions through critical temperatures, in which the sweep in a parameter is costly. To this end, we consider that it is important to take into account the memory and rate effects described here.

\begin{acknowledgments}
This work has been supported by the Spanish State Research Agency (AEI) and the European Regional Development Fund (ERDF, EU) under Project No.~PID2019-105554GB-I00 (MCIN/AEI/10.13039/501100011033). The work of S.Y. was supported by the German
Research Foundation DFG, Project No. 411803875.
\end{acknowledgments}

%\bibliography{btipping}

%

\end{document}